# Common persistence in conditional variance: A reconsideration

chang-shuai Li

*College of Management, University of Shanghai for Science and Technology, Shanghai, 200093, China*

E-mail：chshuai865@163.com

This paper demonstrates the flaws of co-persistence theory proposed by Bollerslev and Engle (1993) which cause the theory can hardly be applied. With the introduction of the half-life of decay coefficient as the measure of the persistence, and both the weak definition of persistence and co-persistence in variance, this study attempts to solve the problems by using exhaustive search algorithm for obtaining co-persistent vector. In addition, this method is illustrated to research the co-persistence of stock return volatility in 10 European countries.

# 1 INTROCUTION

The persistence of financial volatility implies that the shocks of the current conditional variance affects the conditional variances of all future horizons permanently, i.e., the current conditional variance would never diminish. The persistence in the autoregressive conditional heteroscedasticity process means there exists a unit root, which is the so-called IGARCH model proposed by Engle and Bollerslev (1986). The existence of persistence increases the uncertainty of future investment. Fortunately in certain situations investors can eliminate persistence by allocating portfolio to be co-persistent. According to portfolio theory, diversification of assets can reduce risk but meanwhile, it tends to increase the volatility persistence of portfolio which is discussed by Karnasos *et al* (1999). Therefore it is necessary to research volatility co-persistence. Bollerslev and Engle (1993) extend the analysis of persistence to the multivariate vector GARCH process and initiate the research field of the theory of co-persistence in conditional variances.

There are some studies focusing on the theoretical expansion of co-persistence in recent years. Li and Zhang (2001) define the persistence and common persistence of vector GARCH process from the view of the integration and give the properties and the error correction model of vector GARCH process under the condition of the co-persistence. Li and Zhang (2002) study the persistence and co-persistence in SV model instead of GARCH model and present the co-persistence theorem. LIU Dan-hong, XU Zheng-guo and ZHANG Shi-ying (2002) give out the definition of the nonlinear common persistence and make use of the wavelet neural network to approach the nonlinear common persistence then testify the nonlinear common persistence of shanghai and Shenzhen stock market. Du and Zhang (2003) generalize the concept to partial-co-persistence and blocked-co-persistence, they also show the no-existence of co-persistence under two conditions, and existence of co-persistence when two variables exist linear relation. Li and Zhang (2001) discuss the integration and persistence of the BEKK model developed by Engle and Kroner (1995) and suggest the sufficient and necessary condition of co-persistence in variance of the BEKK model. Based on impulse response analysis, Xu and Zhang (2005) puts forward the definition of volatility persistence and common persistence in fractional dimension, and investigates the persistence of FIGARCH process.

It is regrettable that few papers about application study of co-persistence appeared since Bollerslev and Engle (1993) had introduced the co-persistence theory, one of the reasons account for this phenomenon is that the vector GARCH model for analyzing co-persistence has some unavoidable flaws.

This article demonstrates four flaws of analyzing co-persistence by means of vector GARCH. As a result, the co-persistence theory can hardly be applied in the financial field. With the introduction of the half-life of decay coefficient as the measure of the persistence, and the weak definition of both persistence and co-persistence in variance, this study therefore attempts to solve the problem of obtaining co-persistence vector with exhaustive search algorithm. In addition, this method is illustrated to research the co-persistence of stock return volatility in 10 European countries in three financial regions. The result shows that there are seven countries' stocks are co-persistent with others, besides, the stocks in the Germanic area and the Scandinavian area are volatility co-persistent, but the stocks in the French area are not co-persistent.

The rest of the paper is organized as follows: The next section reviews the relative definitions and theorems introduced by Bollerslev and Engle (1993). Section 3 explores the four flaws of co-persistence theory. Section 4 takes the half-life of decay coefficient as the measure of the persistence then put forward the weak definition of both persistence and co-persistence in variance, finally proposes exhaustive search algorithm for obtaining co-persistent vector. Section 5 presents data and empirical analysis of the flaws of the vector GARCH method and the advantages of our method. Section 6 concludes.

## 2  REVIEW OF THE CO-PERSISTENCE THOERY

This section reviews the relative definitions and the corresponding theorems introduced by Bollerslev and Engle (1993).

Let $\{y_t\}$ denote the $N \times 1$ vector stochastic process with the conditional mean and variance functions:

$$E_{t-1}[y_t] = M_t \qquad (2.1)$$

$$\text{var}_{t-1}[y_t] = H_t \quad t=1,2,\ldots \qquad (2.2)$$

$E_{t-1}[\bullet]$ and $\text{var}_{t-1}[\bullet]$ denote the conditional expectation and conditional variance based on the available information set at time $t-1$ respectively. The stochastic $N \times N$ symmetric matrix $H_t$ is almost surely positive definite for all $t$.

Let $\varepsilon_t = y_t - M_t$, and it obeys conditional multivariate normal distribution $N(0, H_t)$. The persistence is well characterized by the influence of the initial conditions on the future conditional variances as the forecast horizon increases. For better illustrating the notion of persistence of a process Engle and Bollerslev gave the following notation:

$$H_t^*(s) \equiv E_s(vech(H_t)) - E_0(vech(H_t)), \qquad t > s > 0 \qquad (2.3)$$

Where $vech(\bullet)$ denotes the vector half operator that stacks the lower triangular elements of an $N \times N$ matrix as an $N(N+1)/2 \times 1$ vector.

**Definition 1**: The stochastic $\{y_t\}$ is defined to be persistent in variance if $\limsup_{t \to \infty} |\{H_t^*(s)\}_i| \neq 0$, a.s. for some s > 0 and some i=1, 2,…, $N(N+1)/2 \times 1$.

In order to research co-persistence among several time series, consider the vector GARCH (p, q) model introduced by Bollerslev, Engle, and Wooldridge (Bollerslev *et al*, 1988).

$$Vech(H_t) = W + \sum_{i=1}^{p} A_i Vech(\varepsilon_{t-i}\varepsilon_{t-i}^T) + \sum_{j=1}^{q} B_j Vech(H_{t-j}) \qquad t=1,2\ldots \qquad (2.4)$$

Conditions on $A_i (i=1,\ldots,p)$ and $B_j (j=1,\ldots,q)$ for $H_t$ to be positive definite a.s.

**Theorem 1**: the vector GARCH (p, q) process $\{\varepsilon_t\}$ defined in (2.4) is covariance stationary if and only if, all the roots of the characteristic polynomial,

$$\det[I - A(\lambda^{-1}) - B(\lambda^{-1})] = 0 \qquad (2.5)$$

Lies inside the unit circle, in which case $\limsup_{t \to \infty} |\{H_t^*(s)\}_i| = 0$, a.s. for all $s > 0$.

Many of empirical results showed that the sum of coefficients of the univariate GARCH (p, q) model

$$\sigma_t^2 = w + \sum_{i=1}^{p} a_i \varepsilon_{t-i}^2 + \sum_{j=1}^{q} b_j \sigma_{t-j}^2 \qquad t=1,2\ldots \qquad (2.6)$$

are often to be very close to one, so the IGARCH (p, q) emerged in response to the needs of times. The shocks to the conditional variance will have a permanent effect as $\limsup_{t \to \infty} |\{H_t^*(s)\}_i| \neq 0$, a.s. for some $s > 0$. This theorem extended the analysis of persistence with the univariate IGARCH (p, q) model to the multivariate GARCH process with unit characteristic root ($|\lambda|=1$).

**Definition 2**: The multivariate stochastic process $\{y_t\}$ is defined to be co-persistent in variance if there exist a nonzero vector $\gamma \in R^N$ such that $\{vec2(\gamma)\}_i \neq 0$ and $\limsup_{t \to \infty} |\{H_t^*(s)\}_i| \neq 0$, a.s. for some $s > 0$ and some i=1, 2,..., $N(N+1)/2 \times 1$, while

$$\limsup_{t \to \infty} |E_s[\gamma^T H_t \gamma] - E_0[\gamma^T H_t \gamma]| = \limsup_{t \to \infty} |Vec2(\gamma^T) H_t^*(s)| \neq 0, \quad a.s.$$

for all $s > 0$. Where $vec2(\gamma) = vech(2\gamma\gamma^T) - diag(\gamma)diag(\gamma)$.

**Theorem 2**: Let $|\lambda_1| \geq |\lambda_2| \geq \cdots \geq |\lambda_r| \geq 1 > |\lambda_{r+1}| \geq \cdots \geq |\lambda_n|$ denote the ordered roots from the characteristic polynomial for the vector GARCH (p, q) process in (2.4), and $v_1, v_2, \cdots v_n$ the corresponding $N(N+1)/2 \times 1$ right eigenvectors,

$$A(\lambda_i^{-1})v_i + B(\lambda_i^{-1})v_i = v_i \tag{2.7}$$

The process is then co-persistence, if and only if

$$vec2(\gamma)'v_i = 0 \qquad (i = 1,...,r) \tag{2.8}$$

for some nonzero vector $\gamma \in R^N$.

**Lemma 1**: linear combinations, $\{\gamma'\varepsilon_t\}$ of the vector GARCH (p, q) process in (2.4) will follow a univariate GARCH (p, q) process if and only if for some scalar constants $\alpha_i, \beta_j$, $i = 1,2...,q, j = 1,...,p$;

$Vec2(\gamma)'A_i = \alpha_i Vec2(\gamma)'$     $i = 1,2,...,q$

$Vec2(\gamma)'B_j = \beta_j Vec2(\gamma)'$     $j = 1,2,...,p$

(2.9)

It clearly shows that if the vector GARCH (p, q) process is co-persistent in variance, and $\{\gamma'\varepsilon_t\}$ follows a univariate GARCH (p, q) model, where $\gamma$ is a co-persistent vector, the sum of the scalar parameters $\alpha_i, \beta_j, i = 1, 2..., q, j = 1, ..., p$; must be less than one.

## 3  THE FLAWS OF VECTOR GARCH METHOD FOR ANALYZING CO-PERSISTENCE

The vector GARCH method for obtaining co-persistent vector reviewed in section 2 has four flaws:

Firstly, the curse of dimensionality problem may arise when estimating vector GARCH model, the unrestricted vector GARCH in (2.4) involves a total of $[2N(N+1) + N^2(N+1)^2(p+q)]/4$ unique parameters, estimator may fail to converge or converge locally.

Secondly, the proof of theorem 2 restricts the eigenvalue of process (2.4) to be real number, however, the eigenvalue is often found to be imaginary number. So the vector GARCH model would fail to analyze the co-persistence property when the eigenvalues contain imaginary number.

Thirdly, Lemma 1 imposes over-identifying restrictions on vector GARCH model (2.4). Let $p = q = 1$ in (2.9) for ease of exposition then it reduces to

$$vec2(\gamma)'A = \alpha vec2(\gamma)'$$
$$vec2(\gamma)'B = \beta vec2(\gamma)' \tag{3.1}$$

Where nonzero vector $\gamma \in R^N$ and scalar $\alpha, \beta \in R$. It easily seen that $vec2(\gamma)'$ is a left eigenvector of coefficient matrix $A$ and $B$ whose corresponding eigenvalue is $\alpha$ and $\beta$. In fact, equation (3.1) makes vector GARCH model (2.4) to be over-restricted with $2\times n$ restrictions which could bring with huge estimator error of co-persistence vector $\gamma$.

Besides, considering conditions on $A$ and $B$ for $H_t$ to be positive definite, the estimation must be biased.

Fourthly, this flaw is most concealed and is the key to the problem. Even if there were no the previous three flaws this defect is enough to make the theory hardly be applicable. We illustrate the problem with one two-dimensional stochastic process and the case of multi-dimension could be deduced similarly.

Given that both $\{y_{1t}\}$ and $\{y_{2t}\}$ follow a non-stationary GARCH model and they are co-persistent, so there exists at least one co-persistence vector $\gamma = (\gamma_1, \gamma_2)$ make the linear combination $\{\gamma_1 * y_{1t} + \gamma_2 * y_{2t}\}$ follow a covariance-stationary univariate GARCH model according to lemma 1. And now we try to obtain $(\gamma_1, \gamma_2)$ by the co-persistence theory, for simplicity we standardize the first fraction to 1 then it becomes into $\gamma = (1, \gamma_2)$.

It is evident that the number of co-persistence vectors often more than one. It should be a set if it exists. But the vector GARCH model with the restriction imposed by lemma 1 estimates the co-persistence vector $\gamma$ by maximum likelihood estimation, which can obtain one and only co-persistence vector $\hat{\gamma}$, which could lead to incorrect conclusion. It is analyzed in detail in the following.

Let $L(\theta) = L(\alpha, \beta, \gamma_2)$ denote maximum likelihood function for estimating the parameter of the vector GARCH model and $\theta$ denote the parameter vector $(\alpha, \beta, \gamma_2)$. The parameter is a three-dimensional vector $\theta = (\alpha, \beta, \gamma_2)$, we can analyze the co-persistence property of $\{y_{1t}, y_{2t}\}$ according to the value of $\hat{\alpha}+\hat{\beta}$ by lemma 1. For better exposition of the problem we reduce three-dimensional parameter vector $(\hat{\alpha}, \hat{\beta}, \hat{\gamma}_2)$ to two-dimensional parameter vector $(\hat{\alpha}+\hat{\beta}, \hat{\gamma}_2)$ so that the maximum likelihood estimation could be illustrated in two-dimensional coordinates graph.

Figure 1. Deriving co-persistence vector $\gamma$ by MLE

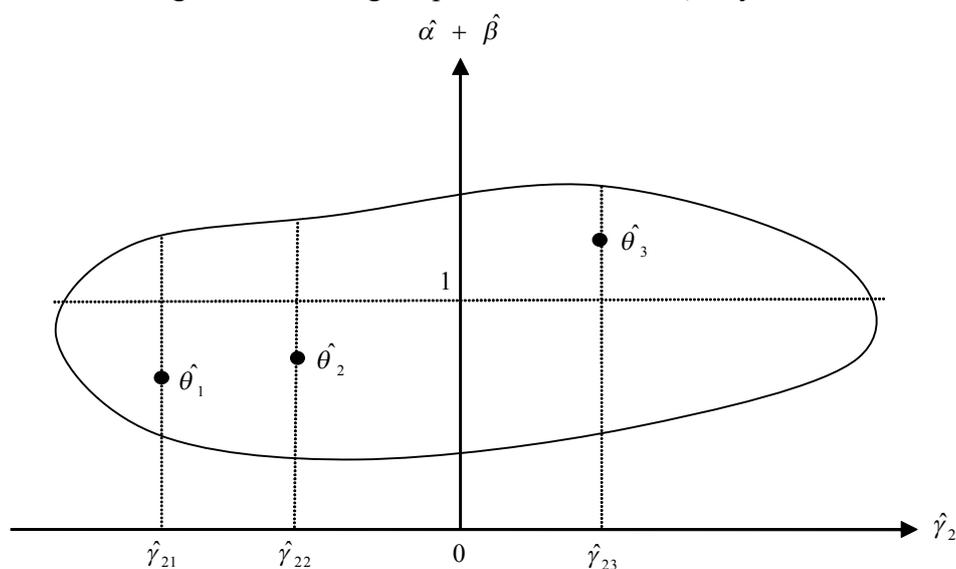

In figure 1, the abscissa axis denotes the co-persistence $\hat{\gamma}$ and the ordinate axis denotes $\hat{\alpha}+\hat{\beta}$. There are more than one likelihood value corresponding to one coordinate point since each value of $\hat{\alpha}+\hat{\beta}$ could be composed of many combinations of $\hat{\alpha}$ and $\hat{\beta}$. For ease of exposition we suppose the closed region is the definition domain of parameter vector $\theta$ in figure 1 (this closed region may differ from the real one, but it would not essentially change the following conclusion). The vector GARCH model selects the MLE $\hat{\theta}$ ($\hat{\alpha}$, $\hat{\beta}$, $\hat{\gamma}_2$) in the closed region by maximizing likelihood function.

Suppose the definition domain of parameter contains only three vectors $\hat{\theta}_1$, $\hat{\theta}_2$ and $\hat{\theta}_3$, which represented by three points in figure 1. These three points lie in the vertical line of the closed region with $\hat{\gamma}_{21}$, $\hat{\gamma}_{22}$ and $\hat{\gamma}_{23}$ respectively. The points $\hat{\theta}_1$ and $\hat{\theta}_2$ are under the horizontal dashed line which means $\hat{\alpha}_i+\hat{\beta}_i<1$ ($i=1, 2$) and the co-persistence vector are $\hat{\gamma}_{21}$, $\hat{\gamma}_{22}$ and the point $\hat{\theta}_3$ is just the opposite. When deriving co-persistence vector $\hat{\gamma}$ by vector GARCH model, we may encounter the following two cases:

Case 1: The likelihood function value reaches maximum at the point $\hat{\theta}_3$. In this case the co-persistence vector $(1, \hat{\gamma}_{21})$ and $(1, \hat{\gamma}_{22})$ can not be obtained, and what can be obtained is $(1, \hat{\gamma}_{23})$ which makes the linear combination of the combination $\{y_{1t}+\gamma_2*y_{2t}\}$ follow an IGARCH model which leads to misjudgement of the co-persistence property of stochastic process $\{y_{1t}, y_{2t}\}$ as it is actually co-persistent with two co-persistence vectors $\hat{\theta}_1$ and $\hat{\theta}_2$.

And what about the probability this case occurs? Each linear combination of the stochastic process $\{y_{1t}, y_{2t}\}$ with the vector of portfolio weight $\gamma$ (it equals to co-persistence vector, the same below) could be fitted by a GARCH model. That is to say, each $\hat{\gamma}_2$ could make $\{y_{1t}+\gamma_2*y_{2t}\}$ follow a GARCH model, even if $\{y_{1t}, y_{2t}\}$ is co-persistent, there certainly exists some $\gamma_2$ make $\{y_{1t}+\gamma_2*y_{2t}\}$ follow an IGARCH model ( consider the extreme case of $\hat{\gamma}_2=0$ and the combination change into the non-stationary $\{y_{1t}\}$ ). So the probability that likelihood function reaches maximum when $\gamma_2=\hat{\gamma}_{23}$ is strict positive.

This case is worst and even if this case did not happen we may still encounter the following case.

Case 2: The likelihood function value reaches maximum at the point $\hat{\theta}_1$. In this case we can only obtain the parameter $\hat{\theta}_1$ ($\hat{\alpha}_1$, $\hat{\beta}_1$, $\hat{\gamma}_{21}$) and only get one of the co-persistence vector $(1, \hat{\gamma}_{21})$, the other co-persistence vector $(1, \hat{\gamma}_{22})$ is missed out. This case is imperfect even it is not worse than case 1.

If we estimate all of the parameters with every $\gamma$ in the domain and then choose the vector of portfolio weight with $\hat{\alpha}+\hat{\beta}<1$. That means we try all the portfolio weight $\hat{\gamma}_{21}$, $\hat{\gamma}_{22}$ and $\hat{\gamma}_{23}$ and then obtain $(\hat{\alpha}_1, \hat{\beta}_1)$, $(\hat{\alpha}_2, \hat{\beta}_2)$ and $(\hat{\alpha}_3, \hat{\beta}_3)$, we can obtain catch both $\hat{\gamma}_{21}$ and $\hat{\gamma}_{22}$, there is no other way available. This is just the idea of exhaustive search algorithm which estimate all of the parameters and then obtain the co-persistence vector by the standard $\hat{\alpha}+\hat{\beta}<1$.

## 4 EXHAUSTIVE SEARCH ALGORITHM

Let's view the notion of decay coefficient $\sum_{i=1}^{p}\alpha_i + \sum_{j=1}^{q}\beta_j$ and half-life K, consider GARCH(1, 1) model

$$\sigma_t^2 = w + \alpha \varepsilon_t^2 + \beta \sigma_{t-1}^2 \qquad t = 1, 2... \tag{4.1}$$

And the $s$ steps expectation of conditional variance is

$$E_t(\sigma_{t+s}^2) = \omega \frac{1-(\alpha+\beta)^s}{1-(\alpha+\beta)} + (\alpha+\beta)^{s-1}\sigma_{t+1}^2 \tag{4.2}$$

It can be seen clearly that the shock to the volatility of $\varepsilon$ is subject to a exponential decay. Then the parameter $\alpha + \beta$ can be called decay coefficient. A more intuitive characteristic of exponential decay is the time required for the decaying quantity to fall to one half of its initial value. This time is called the half-life.

Definition 1 gives the strong definition of the persistence in variance which requires that the decay coefficient must be very close to one. This 'persistence' is the 'permanent persistence', but there is nearly no shock affects economy permanently. We can just take the persistence as 'long-term persistence' which doesn't mean permanent persistence. The formal notion of 'long memory' could explain this issue (as the long memory GARCH introduced by Zhuanxin Ding and Granger (1996)). So we take the half-life of the decay coefficient as measure of persistence in variance. This way is very intuitive and flexible because the investor could choose the half-life of decay coefficient he can afford according to his own risk preference as critical value. The portfolio volatility possesses long memory property if the half-life of decay coefficient is too long. In that situation, the shocks of the current conditional variance would last a long time. Definition 1 is so strong to applicable to empirical analysis, so we put forward the weak definition of the persistence and the co-persistence in variance:

**Definition 3**: Given that the critical value is K, the stochastic process $\{y_t\}$ (one dimension) is persistent in variance if the half-life of decay coefficient of its volatility model is longer than K.

**Definition 4**: Given that the critical value is K, the stochastic process $\{y_t\}$ (multi-dimension) is co-persistent in variance if all the components of it are persistent in variance and there exists a linear combination of them $\{\gamma' y_t\}$ whose half-life of decay coefficient of its volatility model is shorter than K.

Based on these two weak definitions we put forward the steps for testing co-persistence:

1. Test for the persistence in variance of each component of the stochastic process $\{y_t\}$
2. Test for whether exists a linear combination $\{\gamma' y_t\}$ which is not persistence in variance.

The co-persistence test requires all the components of stochastic process are persistent in variance, if there are some components of the stochastic process are not persistent in variance we can only test co-persistence for the rest components, which are entirely in analogy to co-integration test discussed by Engle and Granger (1987).

Now we introduce exhaustive search algorithm. In fact, we can bypassing vector GARCH model which can avoid the four flaws we elaborated above and analyze the persistence of all their linear combinations then choose the ones are not persistent in variance directly, that is the advantage.

As pointed out in the fourth flaw in section 2, exhaustive search algorithm must take all the portfolio weights vector $\gamma$ in their definition domain into consideration, then analyze the persistence of $\{\gamma' y_t\}$ and choose the $\gamma$ make $\{\gamma' y_t\}$ follow a covariance-stationary GARCH model. That is why it is named exhaustive search algorithm. We don't use vector GARCH model in the exhaustive search algorithm so the four flaws are avoided.

Considering $N$ dimensional stochastic process $\{y_t\}$, whose all components $\{y_{it}\}$ are persistence in variance, we need to judge whether it is co-persistent in variance. According to the idea above, we fix portfolio weights $\gamma = (\gamma_1, ..., \gamma_N)'$ first and analyze the co-persistence of the linear combination $\{\gamma' y_t\} = \{\gamma_1 y_{1t} + \gamma_2 y_{2t} + \cdots + \gamma_N y_{Nt}\}$, finally estimate the univariate GARCH model for linear combination. Generally speaking, GARCH (1, 1) model could nearly fit all financial time series very well, so it is reasonable to choose GARCH (1, 1) model to fit the linear combination as most scholars do. After estimating the parameters $\hat{\alpha}, \hat{\beta}$ of GARCH model by maximum likelihood estimation, the mapping can be obtained $f: \hat{\gamma} \to \hat{\alpha} + \hat{\beta}$. Then search all the co-persistence vectors which corresponding decay coefficient $\hat{\alpha} + \hat{\beta} < (\alpha + \beta)_k$ whose half-life is k, that is co-persistence vector set we need.

The case of two-dimensions is shown in figure 2 and the dashed curve line denote the mapping. In the fourth flaw we reduce the this dashed line into just three points $\hat{\theta}_1, \hat{\theta}_2$ and $\hat{\theta}_3$ for easy exposition. The dashed line below horizontal line $(\alpha + \beta)_k$ is the co-persistence vector set we need.

Figure 2. Exhaustive search algorithm

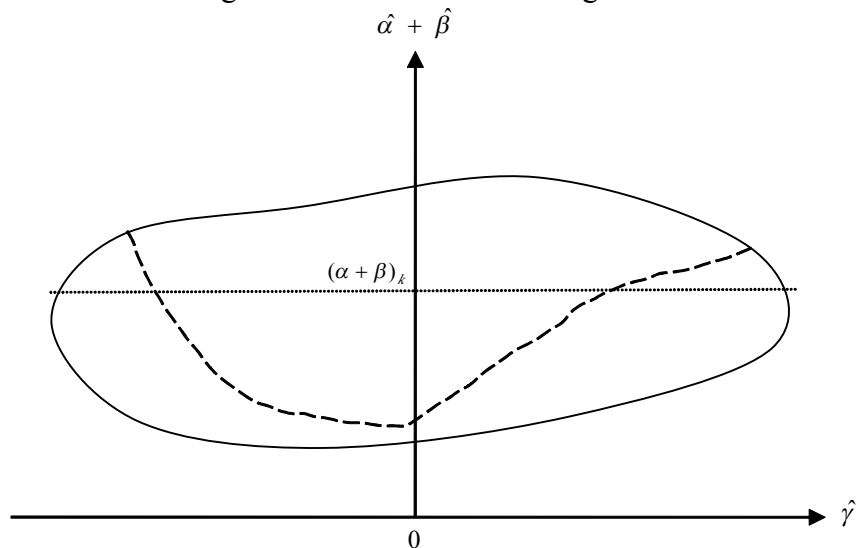

Besides, exhaustive search algorithm can also analyze the co-persistence of any subset of the stochastic process $\{y_t\}$. Let $\gamma^* = (\gamma_1^*, ..., \gamma_N^*)'$ to be the co-persistence vector set obtained, if the component $\gamma_i^*$ of $\gamma^*$ is zero (the number of $\gamma_i$ could be more than only one), then all the subsets of process but $\{y_{it}\}$ are co-persistent. Take a four-dimensional stochastic process $(y_{1t}, y_{2t}, y_{3t}, y_{4t})'$ as example, the co-persistence vectors set contain zero components are $(0, \gamma_2, \gamma_3, \gamma_4)$, $(0, 0, \gamma_3, \gamma_4)$, $(\gamma_1, \gamma_2, \gamma_3, 0)$ then it can be known that $(y_{2t}, y_{3t}, y_{4t})'$, $(y_{3t}, y_{4t})'$, $(y_{1t}, y_{2t}, y_{3t})'$ all are co-persistent.

For simplicity we normalize the first component $\gamma_1$ of $\gamma$ to unity, and then it is only need to consider the rest component $\gamma_i (2 \leq i \leq N) \in (-\infty, +\infty)$. Exhaustive search algorithm searches all the co-persistence vector components $\gamma_i^* (2 \leq i \leq N)$ that make $f(\gamma) = \alpha(\gamma) + \beta(\gamma)$ to be smaller than the given critical value $(\alpha + \beta)_k$. When programming in computer we let the co-persistence vector definition domain to be $\gamma_i (2 \leq i \leq N) \in (-M, +M)$, where $M$ is a very big positive number. The linear combination of the stochastic process can be written as $\{(1/M)y_{1t} + (1/M)\gamma_2 y_{2t} + \cdots + (1/M)\gamma_N y_{Nt}\}$, if $\gamma_i$ don't tend to infinity $\gamma_i / M$ it must tend to zero, so the $i-th$ weight could be neglected. After massive empirical analyses we find the value of $M$ performs well from 5 to 15.

Exhaustive search algorithm is initially used to test the application of co-persistence theory, unavoidably it has disadvantages:

Firstly, it needs huge amount of calculation and the running time would last long if we analyze three or more dimensional stochastic process. But then again the vector GARCH model also need much running time and can hardly converge when analyze a more dimensional stochastic process.

Secondly, the search step may be not small enough to catch all the co-persistence vector component $\gamma_i (2 \leq i \leq N)$, there may exit some singular points. But it doesn't make sense in application because investor can not master these singular points as well.

It is deduced clearly that there is no other effective way but exhaustive search algorithm to overcome the inherent disadvantages of the method proposed by Bollerslev and Engle (1993). Anyhow, exhaustive search algorithm is the suitable method to test the applicability of co-persistence theory and help us to reconsider it.

## 5  DATA AND EMPIRICAL ANALYSIS

### 5.1  Data

In order to illustrate the limitation of the vector GARCH method for obtaining the co-persistence vector this paper uses some countries' stock index in Europe to test whether they are co-persistent and then to know their degree of economic integration. In view of the troubles come from day-of-the-week effects and non-synchronous trading in daily financial time series, and considering a very wide time span may failure to capture the information content of changes in levels and returns. The data are therefore sampled weekly. These different stock indices' data are collected from site: http://finance.yahoo.com. Since the vector GARCH model involves a large number of parameters, the data sample is chosen to be enough long in order to minimize small sample issues. The sample period ranges from February 12, 2001 to August 30, 2010 and the data include 499 observations all together. These data include 10 Europe countries (French, Germany, English, Switzerland, Holland, Austria, Belgium, Danish, Sweden and Norway). Some other countries are not considered in the estimations because of the data unavailability. But these countries are representative so their data are enough to explain the issue. The Europe countries' corresponding stock indices are CAC 40 (French), DAX (Germany), FTSE 100 (English), SMI (Switzerland), AEX (Holland), ATX (Austria), EURONEXT BEL-20 (Belgium), OMXC20.CO (Danish), OMX Stockholm PI (Sweden), OSLO EXCH ALL SHARE (Norway). Then compose weekly total stock returns $R_{i,t} = \log(P_{i,t} / P_{i,t-1})$, where $R_{i,t}$ denotes the continuously compounded return for index $i$ at time $t$, and $P_{i,t}$ denotes the price level of index $i$ at time $t$.

Figure 3. Graphs for each country's stock return series

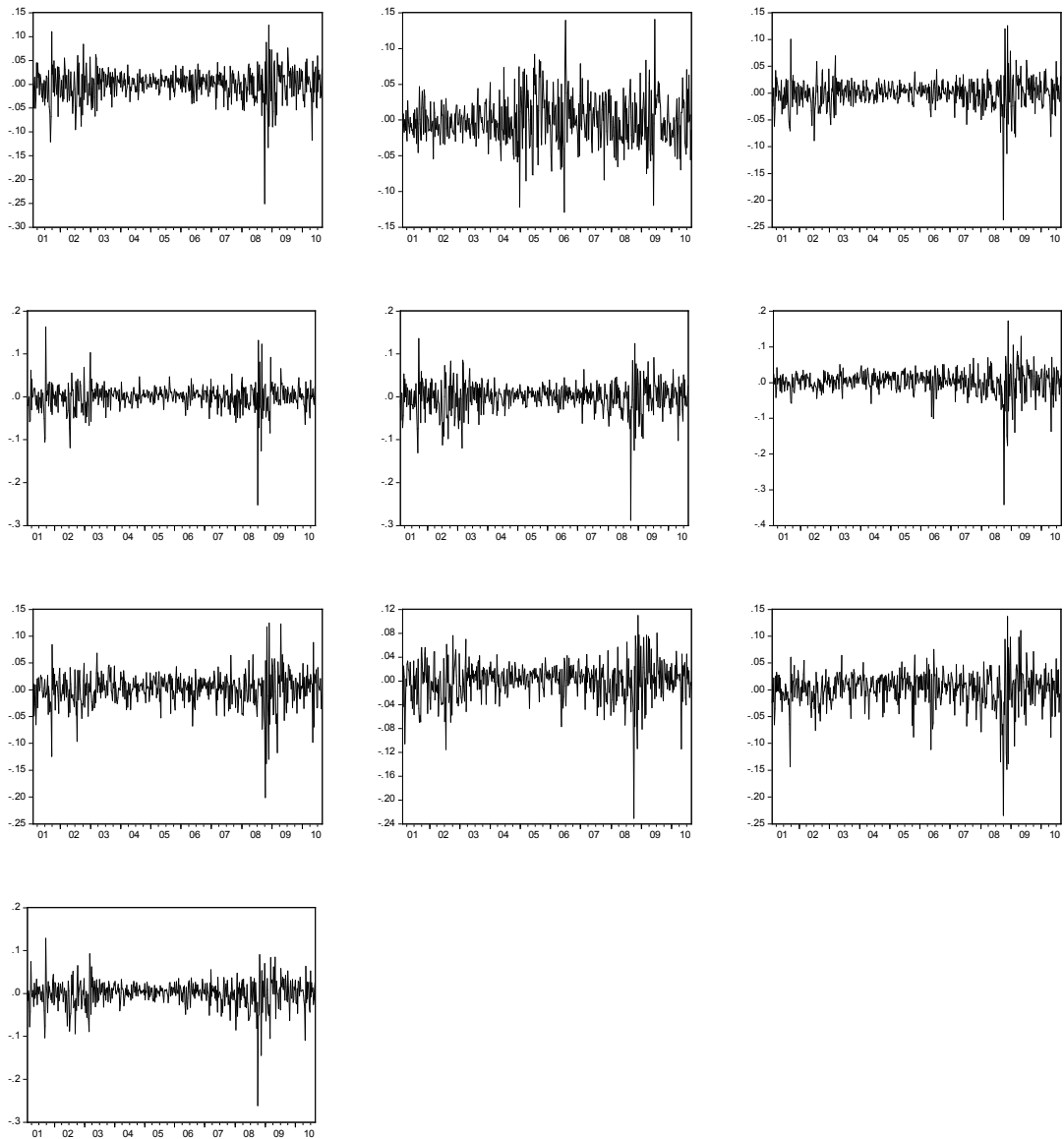

Table 1. Descriptive statistics for Europe countries' stock indices

|             | Mean      | Std. Dev. | Skewness  | Kurtosis  |
|-------------|-----------|-----------|-----------|-----------|
| French      | -0.00091  | 0.033094  | -1.10824  | 10.28806  |
| Germany     | -0.00105  | 0.034092  | 0.232223  | 4.496101  |
| English     | -0.00028  | 0.02752   | -1.20305  | 15.64812  |
| Switzerland | -0.00045  | 0.030131  | -0.98158  | 16.19941  |
| Holland     | -0.0013   | 0.036305  | -1.20707  | 11.46376  |
| Austria     | 0.001516  | 0.036415  | -2.1063   | 21.22116  |
| Danish      | 0.000818  | 0.031204  | -0.91731  | 9.10741   |
| Sweden      | 0.000236  | 0.031908  | -1.078025 | 9. 083388 |
| Norway      | 0.001425  | 0.03486   | -1.26395  | 9.489619  |
| Belgium     | -0.000339 | 0.031827  | -1.486989 | 13.4887   |

Note: All the return series but Germany is negatively skewed and their Kurtosis all excess 3, that indicate all the countries' stock return series are leptokurtosis and fat-tail.

Table 2. Test for heteroscedasticity

|  | LB(3.1) | LB(20) | Test result |
|---|---|---|---|
| French | 85.738 (0.000) | 106.10 (0.000) | H1 |
| Germany | 147.21 (0.000) | 168.34 (0.000) | H1 |
| English | 92.536 (0.000) | 106.64 (0.000) | H1 |
| Switzerland | 89.819 (0.000) | 90.933 (0.000) | H1 |
| Holland | 52.466 (0.000) | 62.337 (0.000) | H1 |
| Austria | 88.017 (0.000) | 109.53 (0.000) | H1 |
| Danish | 193.83 (0.000) | 218.11 (0.000) | H1 |
| Sweden | 64.963 (0.000) | 88.490 (0.000) | H1 |
| Norway | 191.99 (0.000) | 216.92 (0.000) | H1 |
| Belgium | 46.658 (0.000) | 55.021 (0.000) | H1 |

Note: Data summary statistics for weekly returns data. Ljung–Box (LB) at 10 lag lengths and 20 lag lengths statistics are computed for returns and squared returns. And the P-value is given in the parenthesis under the LB statistics. Test results all are H1 indicate this hypothesis reject the null hypothesis of no heteroscedasticity.

Table 3. The estimation of univariate GARCH (1, 1) model for each stock return

|  | constant | $\hat{\alpha}$ | $\hat{\beta}$ | $\hat{\alpha} + \hat{\beta}$ |
|---|---|---|---|---|
| Austria | 3.70E-05 (0.0073) | 0.204927 (0.0000) | 0.788727 (0.0000) | 0.993654 |
| Belgium | 5.09E-05 (0.0022) | 0.273508 (0.0000) | 0.716547 (0.0000) | 0.990055 |
| Danish | 6.65E-05 (0.0020) | 0.161508 (0.0000) | 0.767890 (0.0000) | 0.929398 |
| English | 2.28E-05 (0.0149) | 0.256387 (0.0000) | 0.750195 (0.0000) | 1.006582 |
| French | 1.13E-05 (0.1715) | 0.134848 (0.0000) | 0.871582 (0.0000) | 1.006430 |
| Germany | 2.68E-05 (0.1047) | 0.143088 (0.0000) | 0.841417 (0.0000) | 0.984505 |
| Holland | 4.93E-05 (0.0090) | 0.232757 (0.0000) | 0.754800 (0.0000) | 0.987557 |
| Norway | 0.00011 (0.0000) | 0.161494 (0.0000) | 0.740991 (0.0000) | 0.902485 |
| Sweden | 2.01E-05 (0.039) | 0.078351 (0.0000) | 0.903453 (0.0000) | 0.981804 |
| Switzerland | 5.60E-05 (0.0006) | 0.355956 (0.0000) | 0.658001 (0.0000) | 1.013957 |

Note: The P-value is given in the parenthesis under the parameters. Only the p-values of the constant in the model of French and Germany stock return is more than 5% which means all parameters are significant but these two constants. Additionally, the decay coefficient $\hat{\alpha}+\hat{\beta}$ is very close to 1 which implies all these series are persistent in variance.

Figure 3 shows each of these countries' stock return which strongly indicate that the return series are stationary. Table 1 presents some descriptive statistics for the return series. As is shown in Table 1, all the return series but Germany is negatively skewed and their Kurtosis all excess 3, that indicate all the countries' stock return series are leptokurtosis and fat-tail. Furthermore, Table 2 shows the Ljung–Box tests which clearly suggest the presence of GARCH effects.

5.2 Empirical analysis

In order to compare vector GARCH method and exhaustive search algorithm, the two methods are implemented to analyze co-persistence below. According to the first step of co-persistence test pointed out in section 4, the estimated univariate GARCH models of each stock return series are listed in table 3:

Table 3. The estimation of univariate GARCH (1, 1) model for each stock return

|  | constant | $\hat{\alpha}$ | $\hat{\beta}$ | $\hat{\alpha}+\hat{\beta}$ |
|---|---|---|---|---|
| Austria | 3.70E-05 (0.0073) | 0.204927 (0.0000) | 0.788727 (0.0000) | 0.993654 |
| Belgium | 5.09E-05 (0.0022) | 0.273508 (0.0000) | 0.716547 (0.0000) | 0.990055 |
| Danish | 6.65E-05 (0.0020) | 0.161508 (0.0000) | 0.767890 (0.0000) | 0.929398 |
| English | 2.28E-05 (0.0149) | 0.256387 (0.0000) | 0.750195 (0.0000) | 1.006582 |
| French | 1.13E-05 (0.1715) | 0.134848 (0.0000) | 0.871582 (0.0000) | 1.006430 |
| Germany | 2.68E-05 (0.1047) | 0.143088 (0.0000) | 0.841417 (0.0000) | 0.984505 |
| Holland | 4.93E-05 (0.0090) | 0.232757 (0.0000) | 0.754800 (0.0000) | 0.987557 |
| Norway | 0.00011 (0.0000) | 0.161494 (0.0000) | 0.740991 (0.0000) | 0.902485 |
| Sweden | 2.01E-05 (0.039) | 0.078351 (0.0000) | 0.903453 (0.0000) | 0.981804 |
| Switzerland | 5.60E-05 (0.0006) | 0.355956 (0.0000) | 0.658001 (0.0000) | 1.013957 |

Note: The P-value is given in the parenthesis under the parameters. Only the p-values of the constant in the model of French and Germany stock return is more than 5% which means all parameters are significant but these two constants. Additionally, the decay coefficient $\hat{\alpha}+\hat{\beta}$ is very close to 1 which implies all these series are persistent in variance.

It is legible in table 3 that most of the parameters of univariate GARCH (1, 1) are significant and all these stock return are persistent in variance. Two different methods are implemented to analyze the co-persistence of these stock returns below.

Firstly, we estimate the bivariate GARCH (1, 1) model of the stock return of

Germany and Switzerland with the constraints imposed by lemma 1. The stock volatility of Germany and Switzerland exhibit apparent persistence in table 3. Their continuously compounded percentage daily rate of stock returns, $y_t = (y_{1t}, y_{2t})'$ is here parameterized as bivariate GARCH(1, 1) model.

$$y_t = \begin{pmatrix} -0.002357 \\ (0.035) \\ 0.002683 \\ (0.000) \end{pmatrix} + \varepsilon_t,$$

$$Vech(H_t) = \begin{pmatrix} 0.000336 \\ (0.000) \\ 0.000111 \\ (0.000) \\ 0.000136 \\ (0.000) \end{pmatrix} + \begin{pmatrix} 0.317442 & 0.024214 & 0.005504 \\ (0.000) & (0.000) & (0.205) \\ -0.036395 & 0.180964 & 0 \\ (0.000) & (0.000) & (-) \\ 0 & 0 & 0.250478 \\ (-) & (-) & (0.000) \end{pmatrix} Vech(\varepsilon_{t-1} \varepsilon_{t-1}^T)$$

$$+ \begin{pmatrix} 0.446963 & 0 & 0 \\ (0.000) & (-) & (-) \\ 0 & 0.446963 & -0.071336 \\ (-) & (0.000) & (0.000) \\ 0 & 0 & 0.587776 \\ (-) & (-) & (0.000) \end{pmatrix} Vech(H_{t-1})$$

$$\hat{\gamma} = (1 \quad \hat{\gamma}_2)' = (1 \quad 1.013191)', \quad \hat{\alpha} = 0.243692, \quad \hat{\beta} = 0.446963.$$

In the preliminary estimation the parameters $\{A_1\}_{23}, \{A_1\}_{31}, \{A_1\}_{32}, \{B_1\}_{12}, \{B_1\}_{13}, \{B_1\}_{21}, \{B_1\}_{31}, \{B_1\}_{32}$ were all found to be small and insignificant and then were set to be zero for ease convergence of the nonlinear optimization algorithm. The results show that the portfolio should follow a univariate GARCH (1, 1) with the parameters $\hat{\alpha} = 0.243692, \hat{\beta} = 0.446963$, however, if investor allocate portfolio with weights vector $\hat{\gamma} = (1 \quad \hat{\gamma}_2)' = (1 \quad 1.013191)'$, the portfolio actually follows a univariate GARCH (1, 1) with the parameters $\hat{\alpha} = 0.1671, \hat{\beta} = 0.7034$. This is because of the previous three flaws. The estimated result differs so much from the real situation that vector GARCH can not analyze co-persistence accurately. Even the estimated result is unbiased we still have to face the difficulty brought by the fourth flaw.

Turning to exhaustive search algorithm. Dividing these ten European countries into three regions, that is, the Germanic area (Germany, Switzerland, Austria), the French area (Belgium, France, Holland), the Scandinavian area (Danish, Sweden, Norway), plus England and taking any two countries as one pair (altogether 45 pairs). There are only 7 pairs are found to be co-persistent. In addition, this algorithm is extended to deal with three-dimensional stochastic process to study the co-persistence of the three internal countries' stock return in each region respectively.

This paper sets the critical value of decay coefficient to be 0.8706 (its corresponding half-life is 5), according the weak definition of persistence in variance (definition 3), they are persistent in variance, i.e., each decay coefficient $\hat{\alpha} + \hat{\beta}$ of the GARCH model is bigger than the given critical value 0.8706.

This article sets $\gamma_1 = 1, \gamma_2 \in (-M, M), M = 20, d\gamma_2 = 0.1$ in the exhaustive search algorithm when analyzing the co-persistence of 45 pairs of stocks' return. The graphs of the decay coefficient function $f(\gamma)=\alpha(\gamma)+\beta(\gamma)$ of the 7 co-persistent pairs of stock return are showed in Figure 4.

Figure 4. Graphs of decay coefficient function $f(\gamma)$ of all portfolios

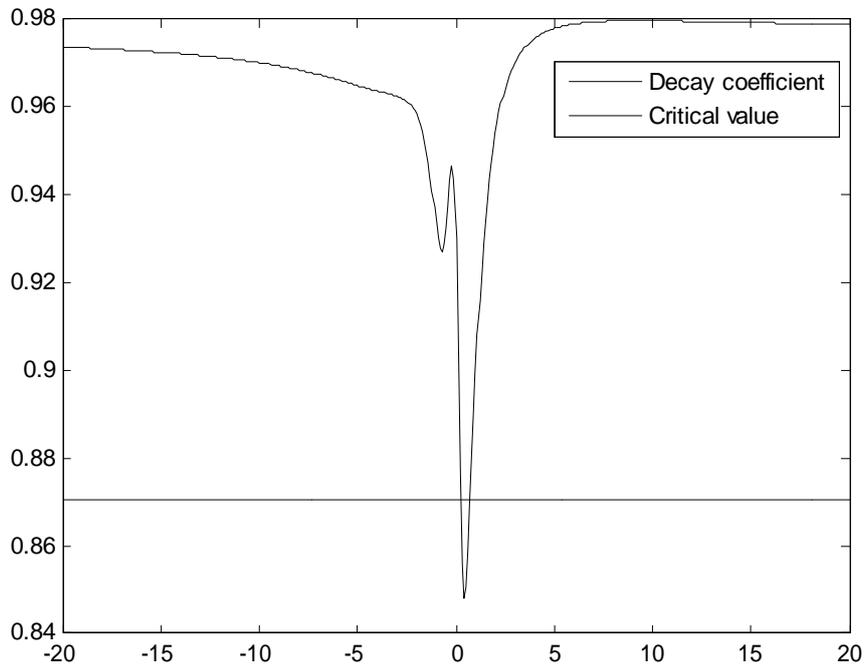

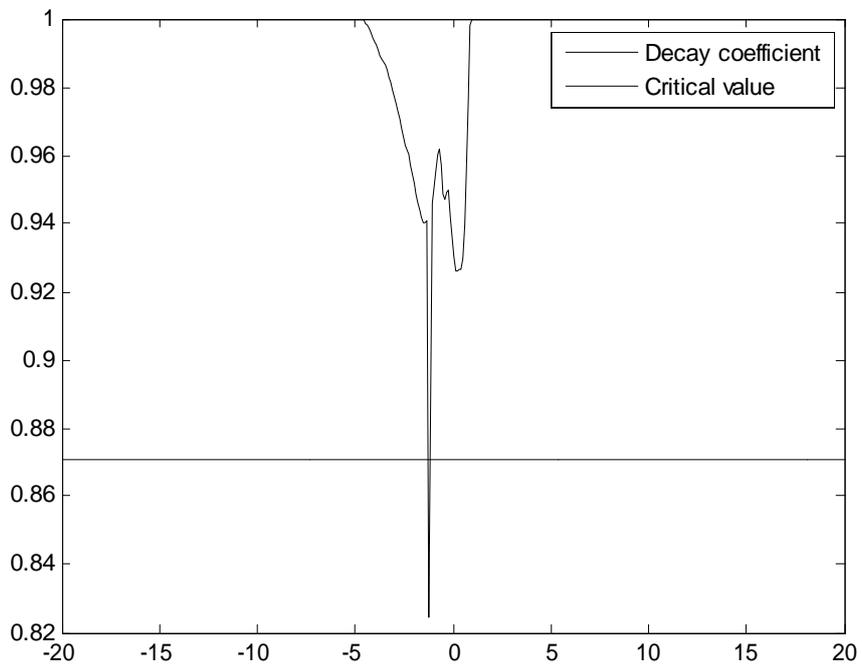

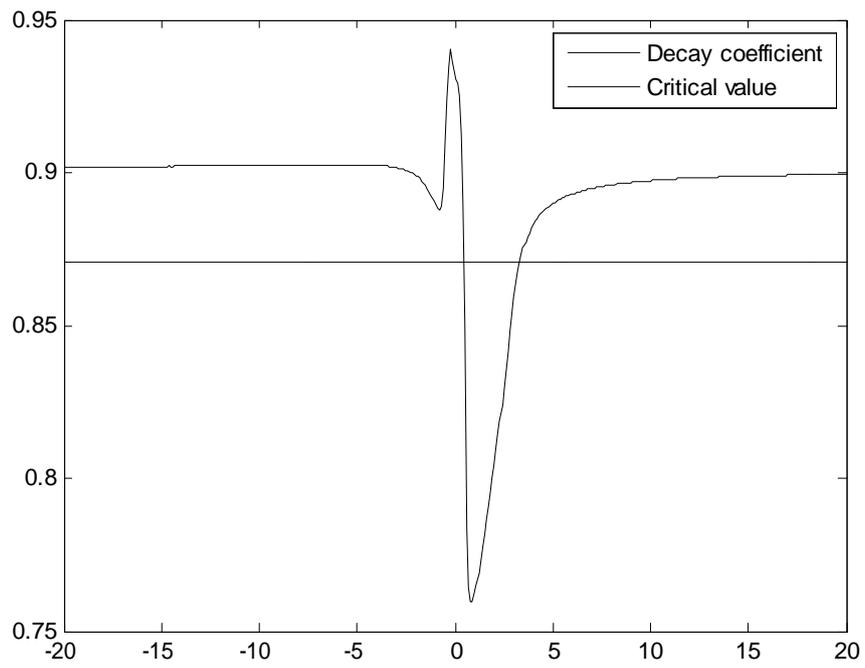

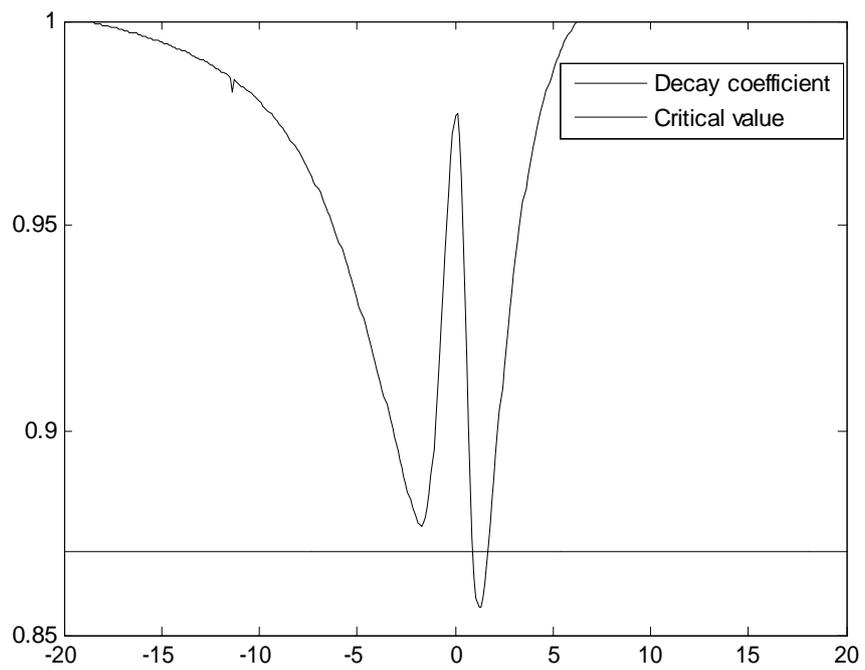

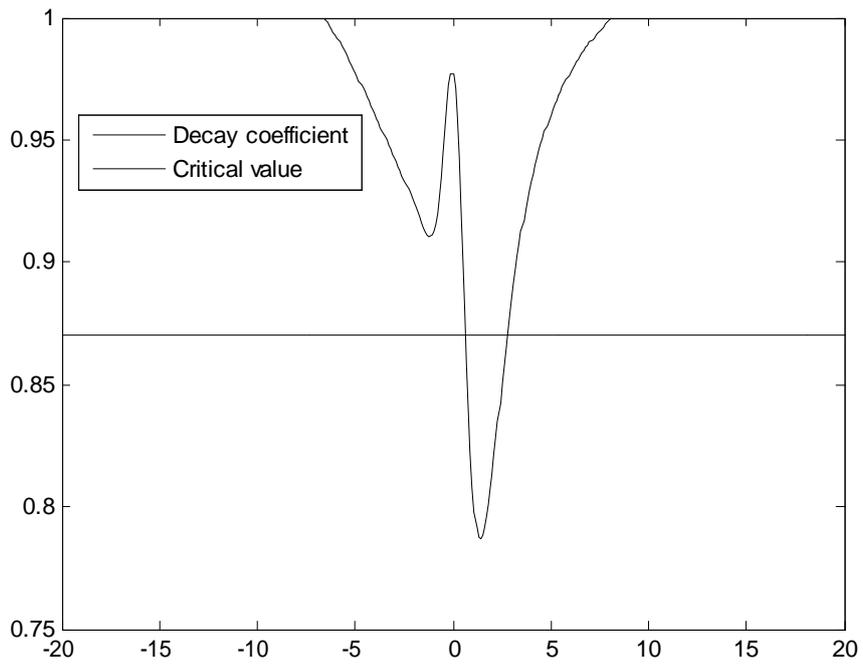

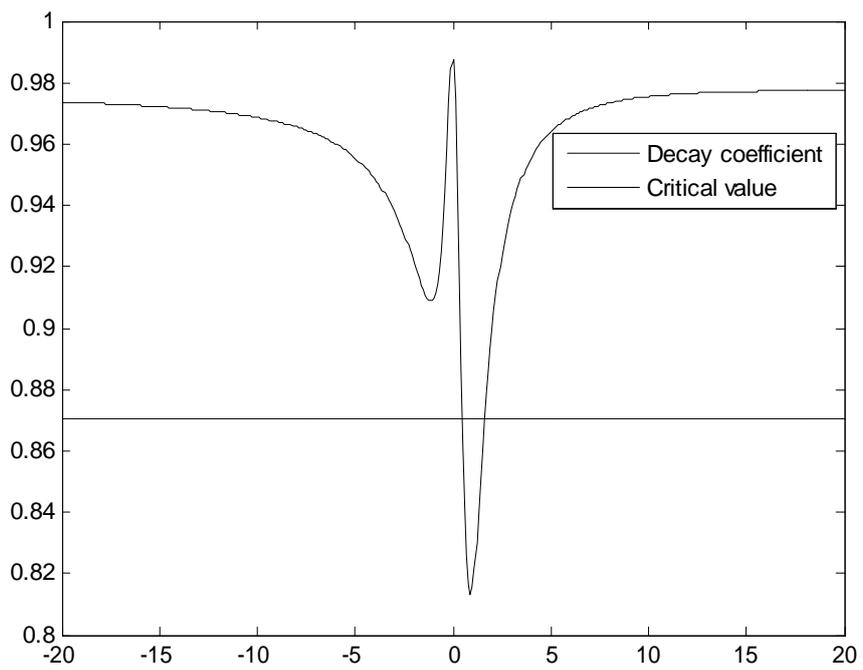

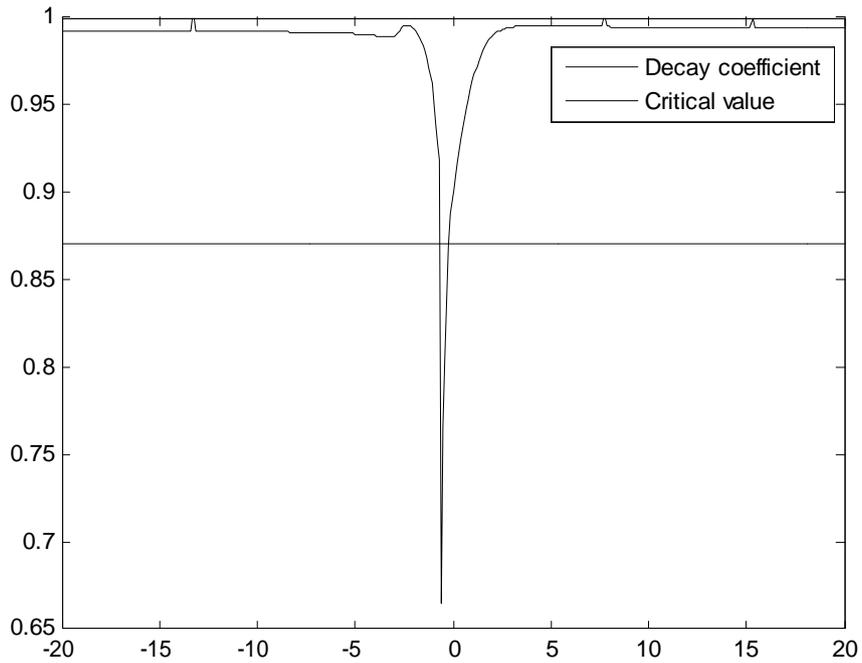

In Figure 4, the ordinate axis and abscissa axis denote decay coefficient and the second component of the vector of portfolio $\gamma_2$ respectively, the dotted lines are the function curves of decay coefficient $f(\gamma)=\alpha(\gamma)+\beta(\gamma)$, and the solid line is the critical value of decay coefficient (0.8706). The intersection of the dotted line with the solid line indicates the two stock returns are co-persistent in variance. The abscissas interval that the curve under the solid line is the co-persistence interval, which means investor can eliminate persistence by allocating portfolio with the co-persistence vector $(1, \gamma_2)$. The results are summarized into table 4.

Table 4. The results of the co-persistence analysis

| portfolio | $(\alpha+\beta)_{min}$ | $\gamma_2^*$ | co-persistence interval |
|---|---|---|---|
| Danish and Germany | 0.8479 | 0.4 | (0.3,0.7) |
| Danish and Switzerland | 0.8245 | -1.2 | -1.2 |
| Danish and Norway | 0.7599 | 0.8 | (0.5,3.3) |
| Germany and Belgium | 0.8576 | 1.2 | (1,1.6) |
| Germany and Switzerland | 0.7870 | 1.4 | (0.7,2.7) |
| Holland and Germany | 0.8133 | 0.9 | (0.5,1.6) |
| Norway and Austria | 0.6650 | -0.6 | (-0.6,-0.2) |

Note: The portfolio can reach the minimum of persistence when $\gamma_2=\gamma_2^*$ with its corresponding coefficient $(\alpha+\beta)_{min}$. And the persistence can be eliminated by allocating portfolio weights vector $(1,\gamma_2)'$, where $\gamma_2$ is inside of the co-persistence interval.

This algorithm is easily extended to analyze multi-dimension stochastic process. This paper studies the co-persistence of the three internal countries' stock return in the each region respectively. Vector GARCH model must encounter curse of dimensionality in three dimensional situation as it have to estimate 78 unique parameters. Set $\gamma_2 \in (-5, 5)$, $\gamma_3 \in (-5, 5)$, and the step $d\gamma_1=d\gamma_2=0.1$, and search the co-persistence vector, we can see Figure 5.

Figure 5. Graph of decay coefficient function $f(\gamma)$ for the Germanic area

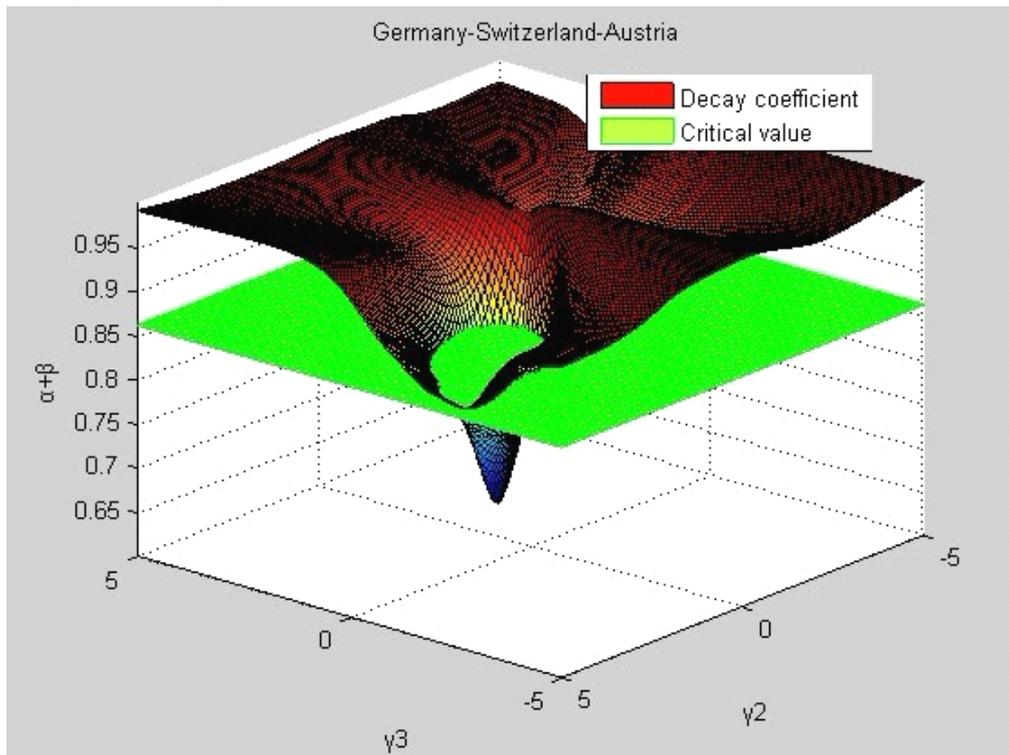

Figure 5 presents the graph of decay coefficient function $f(\gamma)$ for the Germanic area. The axes of coordinates are respectively decay coefficient, the second component of the vector of portfolio $\gamma_2$ and the third component of the vector of portfolio $\gamma_3$. The surface is the function surface of decay coefficient $f(\gamma) = \alpha(\gamma) + \beta(\gamma)$ and the horizontal plane is the critical value of decay coefficient 0.8706(its half-life is 5). The intersection of the surface with the horizontal plane indicates the three stock returns in Germanic area are volatility co-persistent. The area that the surface under the horizontal line is the co-persistence area, which means that investors can eliminate persistence by allocating portfolio with the co-persistence vector $(1, \gamma_2, \gamma_3)'$. Figure 6 is the contour graph of the surface whose altitude is 0.8706:

Figure 6. Contour for the surface of decay coefficient function $f(\gamma)$ for the Germanic area

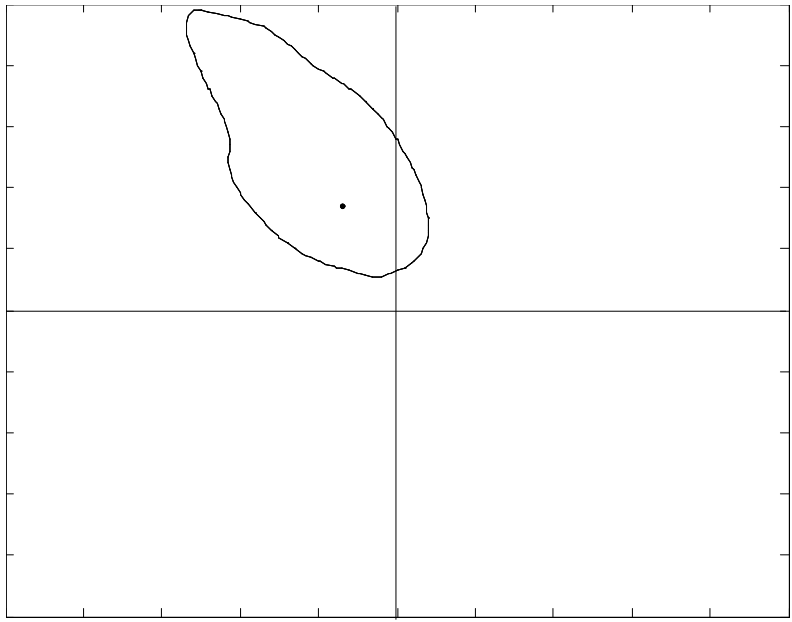

In Figure 6, the loop line is contour line with altitude of 0.8706. Obviously, the Switzerland, Germany and Austria stock return are co-persistent. The area of co-persistence is located inside the closed region. And the portfolio reaches the minimum of persistence in the coordinate point $(\gamma_2, \gamma_3) = (1.7, -0.7)$ with its corresponding coefficient $\hat{\alpha} + \hat{\beta} = 0.6847$. And investor can eliminate persistence by allocating portfolio weights vector $(1, \gamma_2, \gamma_3)'$, where $(\gamma_2, \gamma_3)$ is inside of co-persistence area.

As pointed out in section 4, exhaustive search algorithm can analyze the co-persistence of any subset of the stochastic process. The area of co-persistence contains the axis of coordinates $\gamma_3 = 0$, which implies the co-persistence vector set has zero component, so the first component( Switzerland) and the second component( Germany) of this stochastic process are co-persistent which is in accordance with the two dimensional result ( see Figure 4).

Figure 7. Graph of decay coefficient function $f(\gamma)$ for the French area

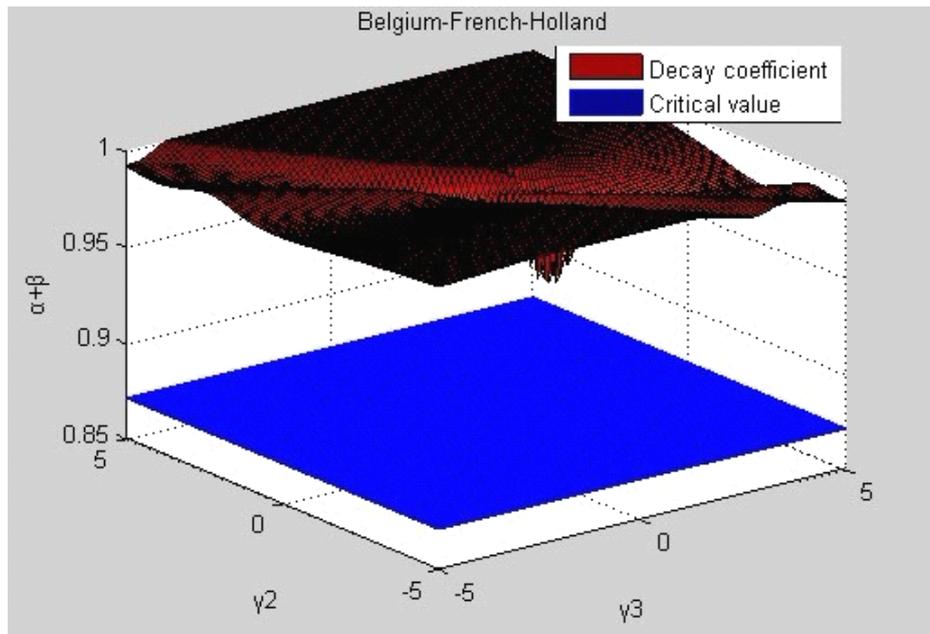

Figure 7 presents the graph of decay coefficient function $f(\gamma)$ for the French area, and the portfolio can reach the minimum of persistence in the coordinate point $(\gamma_2, \gamma_3) = (-1.1, 0.7)$ with its corresponding coefficient $\hat{\alpha} + \hat{\beta} = 0.9427$, but the persistence still can not be eliminated as the decay coefficient function surface is disjoint with the horizontal plane. There is no any subset of this stochastic process is co-persistent which is in accordance with in accordance with the two dimensional result ( see Figure 4).

Figure 8. Graph of decay coefficient function $f(\gamma)$ for the Scandinavian area

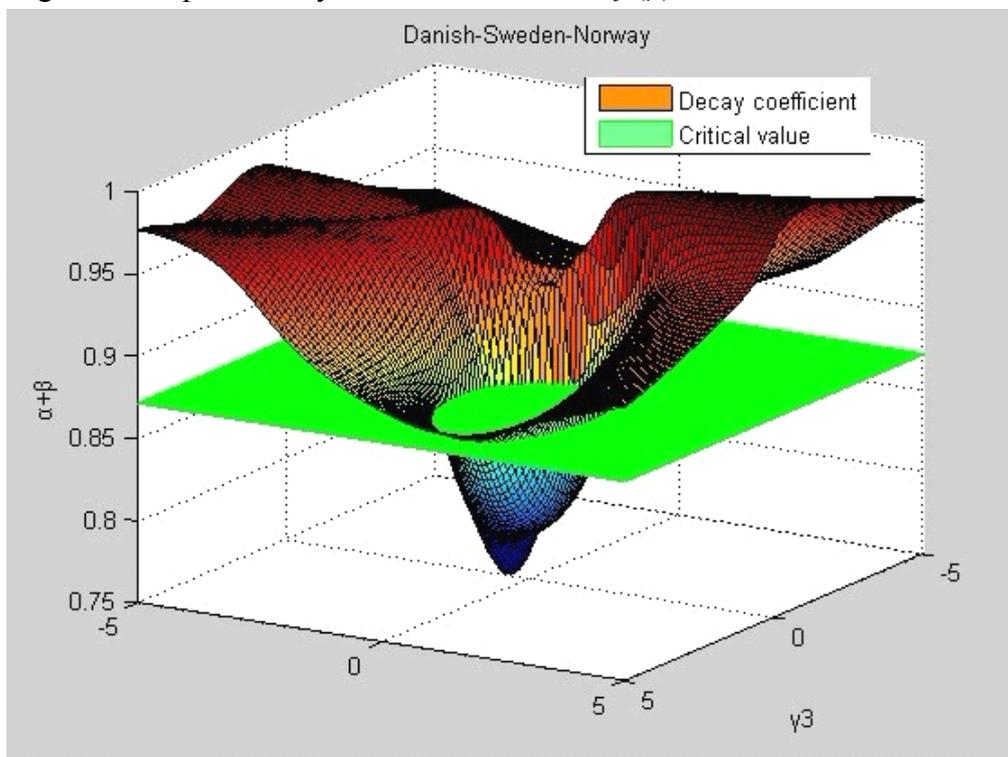

Figure 8 presents the graph of decay coefficient function $f(\gamma)$ for the Scandinavian area, the surface intersects with the horizontal plane, means the Danish, Sweden and Norway stock return are co-persistent.

Figure 9. Contour for the surface of decay coefficient function $f(\gamma)$ for the Scandinavian area

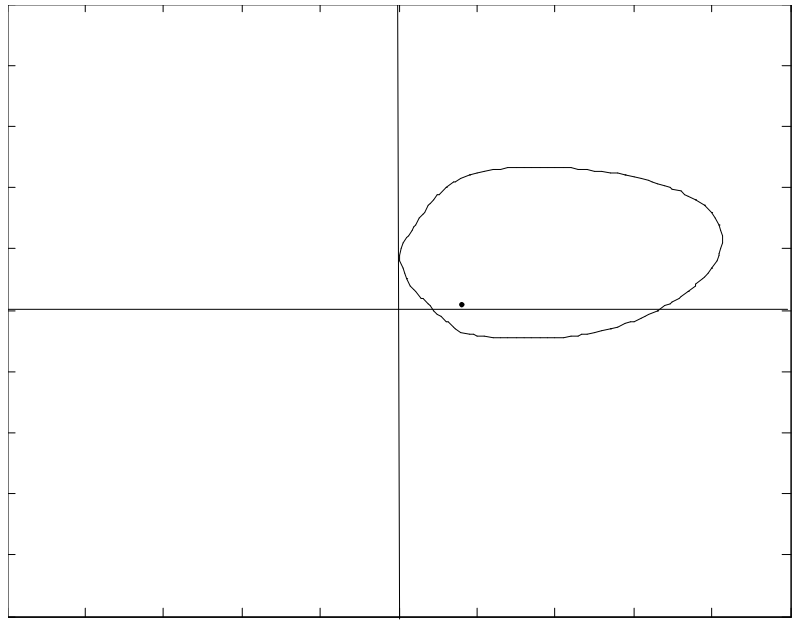

Figure 9 is the contour graph of the surface whose altitude is 0.8706. The loop line is contour line with altitude of 0.8706. The area of co-persistence is located inside the loop line. And the portfolio can reach the minimum of persistence in the coordinate point $(\gamma_2, \gamma_3) = (0.1, 0.8)$ with its corresponding coefficient $\hat{\alpha} + \hat{\beta} = 0.7584$. The persistence can be eliminated by allocating portfolio weights vector $(1, \gamma_2, \gamma_3)'$, where $(\gamma_2, \gamma_3)$ is inside of co-persistence area. The area of co-persistence contains axis of coordinate $\gamma_2 = 0$, which implies the co-persistence vector set has zero component, so the first component( Danish) and the second component( Norway) of this stochastic process are co-persistent which is in accordance with in accordance with the two dimensional result ( see Figure 4).

Figure 10. The comprehensive results of co-persistence in the three regions

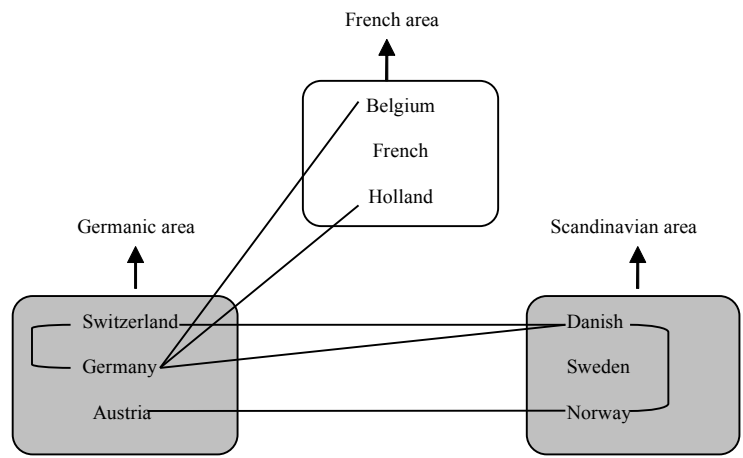

Figure 10 presents the above abundant results in a brief way. In the chart, there are three blocks stand for three regions. The line connects two countries means the stock return of the two countries are co-persistent, and the blocks filled with grey colour means the three stock return of these three countries are co-persistent. There is only one pair of stock return are co-persistent in the Germanic area, and no any pair of stock return are co-persistent in the French area and one pair of stock return are co-persistent the Scandinavian. There are two pairs of stock return are co-persistent between the Germanic area and the French area, three pairs of stock return are co-persistent between the Germanic area and the Scandinavian area and no any pair of stock return are co-persistent between the French area and the Scandinavian area. The stock returns of the countries in the Germanic area and the Scandinavian area are co-persistent but the French area.

Just as its name implies, exhaustive search algorithm may lack of technical content, but what we focus on is to test the actual application of co-persistence by this algorithm. Though this method is simple but the results derived by it offer a huge amount of information. It can be easy realized, easily extended to multi-dimensional and comprehensive analysis of the co-persistence of stochastic process. This algorithm increases the practical applicability of co-persistence theory.

# 6 CONCLUSION

After analyzing the co-persistence theory this paper points out the four flaws of the vector GARCH method for studying co-persistence property. In order to surmount these flaws we take the half-life of decay coefficient as the measure of the persistence and put forward the weak definition of the persistence and the co-persistence in variance. On these bases we use exhaustive search algorithm for obtaining co-persistent vector. In addition to overcome the four flaws it can also analyze the co-persistence of any subset of the stochastic process. This method is illustrated by applying it to study the co-persistence of stock return volatility in 10 European countries, the conclusion is drawn that there are seven countries' stocks are co-persistent with others, and the stocks in the Germanic area and the Scandinavian area are volatility co-persistent, the stocks in the French area are not co-persistent, which also implies the high degree of integration of the economies in Europe.

Our study can be extended in three ways: Firstly, the data of this paper are collected weekly while the high-frequency data are easily to be obtained, so the co-persistence of the stochastic process may be different by using the high-frequency data. Secondly, SV model (stochastic volatility model) has some good features for fitting financial time series, so researchers could use SV model instead of GARCH model to analyze the co-persistence. Thirdly, some empirical results obtained by Lamoureux *et al* (1990) show that structural change in some extent accounts for the persistence in variance, so GARCH model or SV model with structural change discussed by Xu and Zhang (2005) can be taken into consideration for co-persistence analysis.